\documentclass[conference]{IEEEtran}
\IEEEoverridecommandlockouts
\usepackage{cite}
\usepackage{amsmath,amssymb,amsfonts}
\usepackage{algorithmic}
\usepackage{graphicx}
\usepackage{textcomp}
\usepackage{xcolor}
\usepackage{balance}
\usepackage{makecell}
\usepackage{hyperref}
\usepackage{url} 
\usepackage{listings} 
\usepackage{float}
\usepackage[skip=5pt]{caption}
\usepackage{subfig}

\usepackage[letterpaper,%
            left=0.75in,right=0.75in,top=0.75in,bottom=1in,%
            footskip=.25in,bindingoffset=0.2in]{geometry}

\def\BibTeX{{\rm B\kern-.05em{\sc i\kern-.025em b}\kern-.08em
    T\kern-.1667em\lower.7ex\hbox{E}\kern-.125emX}}

\begin{document}

\makeatletter
\newcommand{\linebreakand}{%
 \end{@IEEEauthorhalign}
 \hfill\mbox{}\par
 \mbox{}\hfill\begin{@IEEEauthorhalign}
}

\makeatother

\title{Data Warehouse Design for Multiple Source Forest Inventory Management and Image Processing

} 

\author{

\IEEEauthorblockN{Kristina Cormier}
\IEEEauthorblockA{\textit{Computer Science} \\
\textit{Okanagan College}\\
Kelowna, Canada \\
0009-0004-3783-9704}

\and

\IEEEauthorblockN{Kongwen (Frank) Zhang}
\IEEEauthorblockA{\textit{School of Computing} \\
\textit{University of the Fraser Valley}\\
Abbotsford, Canada \\
0000-0002-8002-5341}

\and

\IEEEauthorblockN{Joshua Padron-Uy}
\IEEEauthorblockA{\textit{Computer Science} \\
\textit{Okanagan College}\\
Kelowna, Canada \\
0009-0004-1038-0833}

\linebreakand

\IEEEauthorblockN{Albert Wong}   \IEEEauthorblockA{\textit{Mathematics and Statistics} 
  \\   \textit{Langara College}\\
   Vancouver, Canada \\
   0000-0002-0669-4352}

 \and
 
   \IEEEauthorblockN{Keona Gagnier}
    \IEEEauthorblockA{\textit{Computer Science} \\
   \textit{Okanagan College}\\
   Kelowna, Canada \\
    0009-0002-9274-6856}

  \and
    
   \IEEEauthorblockN{Ajitesh Parihar}
    \IEEEauthorblockA{\textit{Computer Science} \\
   \textit{Okanagan College}\\
    Kelowna, Canada \\
   0009-0001-3162-1470}

}

\maketitle

\begin{abstract}
This research developed a prototype data warehouse to integrate multi-source forestry data for long-term monitoring, management, and sustainability. The data warehouse is intended to accommodate all types of imagery from various platforms, LiDAR point clouds, survey records, and paper documents, with the capability to transform these datasets into machine learning (ML) and deep learning classification and segmentation models. In this study, we pioneered the integration of unmanned aerial vehicle (UAV) imagery and paper records, testing the merged data on the YOLOv11 model. Paper records improved ground truth, and preliminary results demonstrated notable performance improvements.

This research aims to implement a data warehouse (DW) to manage data for a YOLO (You Only Look Once) model, which identifies objects in images. It does this by integrating advanced data processing pipelines. Data are also stored and easily accessible for future use, including comparing current and historical data to understand growth or declining patterns. In addition, the design is used to optimize resource usage. It also scales easily, not affecting other parts of the data warehouse when adding dimension tables or other fields to the fact table. DW performance and estimations for growing workloads are also explored in this paper.
\end{abstract}


\section{Introduction}

Individual tree bioinformation, such as species, height and diameter at breast height (DBH), forms the foundational basis for effective forestry management \cite{wang2024individual}. Traditional individual tree species classification often relies on limited datasets, typically derived from a single image source  \cite{chen2021new}. Although significant progress has been made by numerous experiments that have been reported, most methodologies are validated under predefined conditions and lack generalizability to other regions \cite{qin2022individual}.

Canada's vast forested areas have generated a tremendous amount of data, much of which remains unprocessed. To address this challenge, scientists from the Canadian Forest Service have reached a consensus on the need for remote sensing techniques capable of covering large areas, coupled with automated, generalizable models for efficient data processing and information retrieval \cite{white2016remote}.

We also recognize the potential advancement in forest management through the use of various datasets collected over time from the same logging sites, sourced from various platforms and in different formats \cite{lausch2018monitoring}. This approach necessitates the development of a new data warehouse capable of integrating multiple data sources and automatically converting them into standardized formats for efficient data processing and information retrieval.

In this study, we began by integrating paper documents by synchronously obtaining the coordinates, which served as both ground truth and a basis for post-processing validation. These highly accurate documents were first used to verify ground-truth data generated through visual interpretation, enhancing the initial classification process. The final results were individually validated against the paper documents for accuracy.

Preliminary findings reveal that visual interpretation struggles in areas with multi-tree crown tops and dense canopies, with machine learning (ML) approaches achieving only 59\% accuracy under these conditions. However, incorporating paper records significantly improved classification accuracy, demonstrating their value in enhancing the reliability of the results. Different solutions are available for YOLO implementations to handle various data-driven environmental problems. However, integrating diverse data sources and performing the necessary transformations to meet classification objectives have some common challenges when handling these data. 

To address these challenges, this project introduces a data warehouse (DW) that uses a star schema and is tailored for individual tree species (ITS) classification. The system leverages YOLO as the core component for data ingestion. The capabilities of YOLO's real-time preprocessing, combined with its ability to detect and classify objects with high accuracy, make it an excellent option for handling complex, large-scale forest data sets. 

The DW serves as a central repository for managing data collected from various sources, such as satellite imagery, aerial photography, and ground-based surveys. The system streamlines data ingestion, storage, and spatial and spectral information analysis for ITS by integrating advanced data processing pipelines. YOLO detects trees and classifies them into species categories during ingestion. The DW ensures that the data is both structured and ready for analysis. 

This is an innovative approach that bridges the gap between advanced machine learning (ML) and practical forestry applications. This enables researchers and stakeholders to access high-quality, actionable insights. The DW contributes to developing more efficient, scalable, and data-driven forestry management strategies by automating key processes and enhancing data accessibility. Furthermore, allowing users from multiple organizations to access a single centralized repository for each set of measurement data ensures consistent data throughout the enterprise.\cite{kimball2013data}

\section{Existing Works} 
The DW in this project incorporates an object detection model called YOLO, a widely used framework for object detection and image segmentation. Initially developed by Joseph Redmon and Ali Farhadi at the University of Washington and first introduced in 2015, YOLO quickly gained recognition for its exceptional speed and accuracy \cite{farhadi2018yolov3} \cite{redmon2017yolo9000}. Although YOLO is not inherently part of the DW architecture, it has been integrated into the pipeline to demonstrate the warehouse's data processing capabilities.

The current study uses YOLOv11, the latest version developed by Ultralytics. This advanced model delivers state-of-the-art (SOTA) performance across various tasks, including detection, segmentation, pose estimation, tracking, and classification. It enhances AI applications across diverse domains. YOLOv11 is licensed under the AGPL-3.0 license for research purposes \cite{yolo11_ultralytics}.

A data warehouse is a system used for reporting and data analysis and is considered a core
component of a business intelligence environment. Due to the varying definitions of this system, it 
is considered as a vital pivotal building block in the decision-making process within an organization. 
Data in the data warehouse system is not frequently updated, therefore smaller organizations may 
not have a data warehouse unless they have other analytical systems available. Data warehouses are 
systems that are designed for query and analysis rather than transaction processing, and they usually 
contain large amounts of historical data \cite{khan2023data}.

In 2024, Parihar \cite{parihar2024algorithmic} describes how this schema design allows for fast and efficient querying, enabling ML models and human analysts to retrieve relevant data in real time. Additionally, the DW is the central repository for all historical data, supporting analytical
tasks and predictive modelling.

Recently, in 2025, Ghadimi \cite{ghadimi2025databay} explains that in order to build a data warehouse as a single source of truth, it is
necessary to gather all data from various data sources and
prepare it for use. The data model must be designed in a way that
allows downstream users to easily access the data with high
performance. Furthermore, depending on the business
requirements, data must be stored with specified granularity in a
historical format.

Modern forest management requires research and analysis, forest inventories and stand growth projection modelling. The accumulated data is shared across government and industry to support informed decision making \cite{bc_forest_inventory}. This supports the need for a centralized information system that is capable of efficient analysis of multiple data-types.

\subsection{Data Warehouse Design}
Our DW design is structured to support identifying and classifying individual tree species using images. It is designed to store raw data. Use cases include storing uploaded data, providing data for YOLO outputs, and analytical results for long-term use. Future work will incorporate other data sources such as text, video, and remote sensors. 

Fig.~\ref{fig:multiple_source} illustrates the star scheme we used for the DW design. It organizes data into one fact table that is surrounded by multiple dimension tables. This design is simple, intuitive, and easy to understand and visualize. In addition, it has a one-to-many relationship between the fact and dimension tables to facilitate straightforward queries, optimize performance, and enable faster query execution. Dimension tables store descriptive information about the entities involved in the business process, such as tree species or image files. These tables are de-normalized for improved performance by limiting complex joins.  The fact table stores quantitative data associated with the entities in the dimension tables. 

This architecture supports complex analyses on various dimensions, such as analyzing metrics by species over time or filtering by image properties. Furthermore, it promotes scalability. New metrics can be added to the fact table. New dimension tables can be incorporated without disrupting the DW structure. This DW architecture also offers flexibility in reporting such as trends over time, image usage by resolution, or performance metrics grouped by conservation status. In short, the DW with a star schema provides an efficient, intuitive, and scalable structure for analyzing metrics by date, image, and species attributes. 

The current DW design has three dimension tables. The Dim-Date table facilitates a structured representation of time. It enables temporal analysis, such as understanding tree health trends or changes in species distribution over time. The Dim-Image table describes details of the images used for tree identification. This supports queries about the images used, their sources, and their technical characteristics. The Dim-Species table stores taxonomic and descriptive information on tree species. This supports analyses of species distribution, diversity, and ecological patterns. As mentioned, the Fact-Tree-Metrics fact table is linked to the dimension tables and stores data on individual trees. This table consolidates all the quantitative data and their relationships with relevant dimensions. Future expansions include dimension tables for text, video, and remote sensor data. 

\begin{figure*}[!t]
\centering
\includegraphics[width=0.8\textwidth]{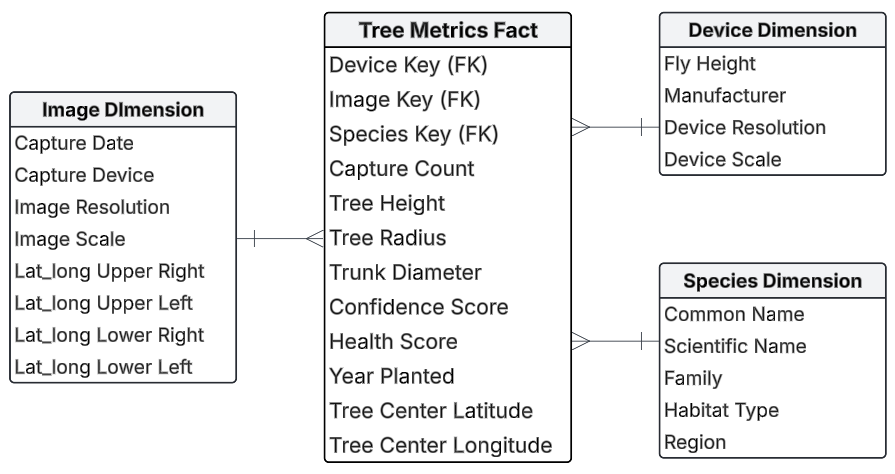}
\caption{DW Design for Individual Tree Species Classification}
\label{fig:multiple_source}
\end{figure*}

\begin{figure*}[!t]
\centering
\includegraphics[width=0.8\textwidth]{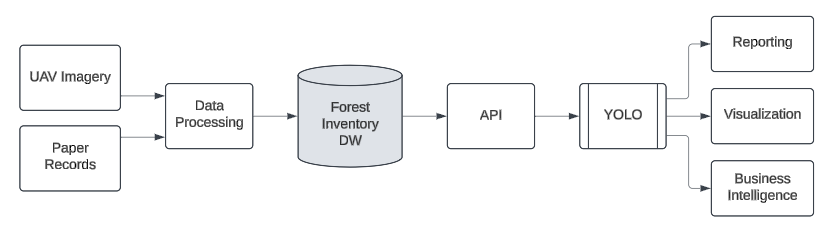}
\caption{Forest Inventory System
Architecture}
\label{fig:system_architecture}
\end{figure*}



\subsection{Data Ingestion and Processing}
YOLO's primary contribution to the DW is the ingestion process that involves visual or image-based data. YOLO improves the DW design in several ways. One is in the data ingestion and metadata generation process. In forestry or environmental monitoring applications, images or videos of forested areas are often captured by drones, satellites, or stationary cameras. YOLO has the ability to process these visual data to identify ITS, classify them, and extract metadata. This includes information such as tree species detected, the count of each species, and the geographical coordinates or location of the detected trees. Images for this process were stored online and used from previous work \cite{roboflow_uav_tree}. An Application Programming Interface (API) is used to connect the DW model and the YOLO model, allowing for data access. These processes are illustrated in Fig.~\ref{fig:system_architecture}. 

In addition, YOLO generates data to improve the alignment of tree species data with environmental metrics. This can include growth rates, carbon sequestration, and biodiversity analysis. This is key data for Natural Resources Canada \cite{nrcan_forest_carbon}.

Furthermore, YOLO allows for real-time event monitoring. Using its real-time capabilities, data streams from drones or satellite images can feed directly into the DW. For example, it can detect and classify trees during a live flyover of a forest and automatically update the DW with real-time species distribution and density data. This is similar to real-time customer behavior analysis in retail. This real-time capability enables on-the-fly decision making for conservation efforts, logging restrictions, or reforestation projects.

Finally, it supports enhanced analytics for forestry management. The extracted and structured data can be queried and analyzed alongside other environmental factors, such as soil quality or deforestation rates. This can include identifying areas with declining biodiversity by comparing historical species data or assessing the impact of natural disasters, such as fires or storms, on tree populations.

YOLO combined with a DW is more equipped to handle diverse data types, enabling richer analytics and faster decision-making.

\section{Estimating the Size of the Data Warehouse}
The existing records with the described pre-processing and augmentations for the research involve four data sets of three consecutive days with two data sets taken for the third day. The four data sets contain 22, 50, 116, and 50 total images. Therefore, the average number of images taken per day is calculated as follows.

\begin{itemize}
    \item Average per day = \( (22 + 50 + ((116 + 50) / 2)) / 3 = 51 \)
\end{itemize}

The Image Dimension is expected to grow at a similar rate to the fact table, with one record in both tables corresponding to a single image. 
 Table \ref{tab:current_data} presents the current size of the DW by getting the sum of all versions.

\begin{itemize}
    \item DW Total Record Count: \( 22 + 50 + 116 + 50 = 238 \)
\end{itemize}

\begin{table}[ht]
	\centering
	\caption{Current DW Size}
	\begin{tabular}{|l|l|l|}
	\hline
	\textbf{Type}& \textbf{No. of Records}& \textbf{MiB}\\
	\hline
	Image Dimension& 238& 920.9\\
	\hline
	Fact Table& 238& 0.01\\\hline
	\end{tabular}
	\label{tab:current_data}
\end{table}

Given that tree classification research aims to support urban planning and management, we assume that the data will grow quarterly every year as follows. 

\begin{itemize}
    \item Yearly Data Growth: \( 51 * 4 = 204 \)
\end{itemize}

Based on the above, the estimated size of the DW size after 10 years described in Table \ref{tab:estimated_data} will require much more storage space. We calculate the estimated 10-year record count by considering the Yearly Data Growth.

\begin{itemize}
    \item Estimated \# of records - 10-years: \( 51 * 4 = 204 \)
\end{itemize}

\begin{table}[ht]
	\centering
	\caption{Estimated Size of the DW - 10 years}
	\begin{tabular}{|l|l|l|}
	\hline
	\textbf{Type}& \textbf{No. of Records}& \textbf{GiB}\\
	\hline
	Image Dimension& 2072& 8.12\\
	\hline
	Fact Table& 2072& 0.000085\\\hline
	\end{tabular}
	\label{tab:estimated_data}
\end{table}

\section{Image Transformed for YOLOv11}
Fig.~\ref{train_val} Illustrates images used for training (a) and the corresponding values (b).

Fig.~\ref{results} Shows the results of YOLOv11.

\begin{figure*}[!t]
\centering
\subfloat[a]{\includegraphics[width=2.5in]{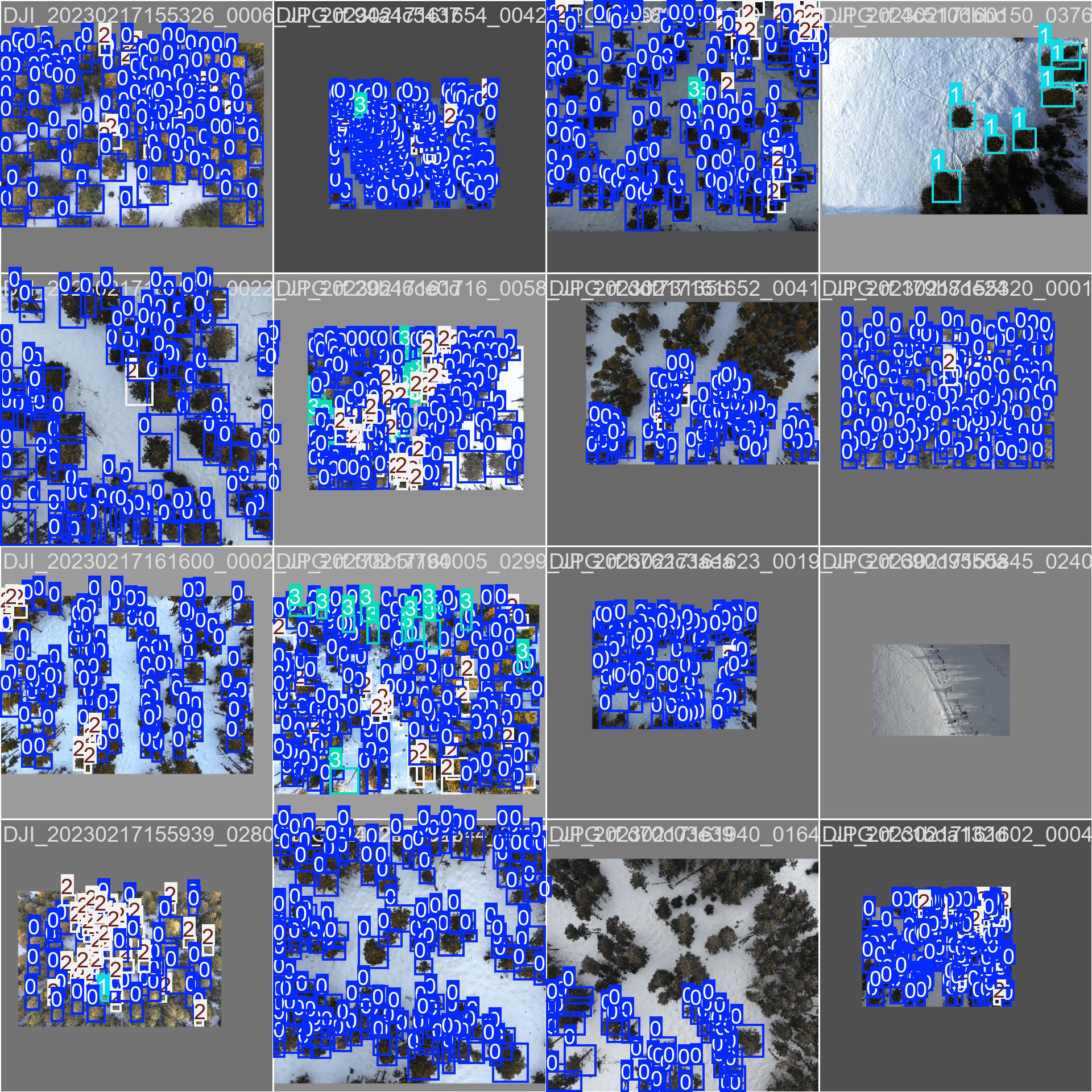}}
\subfloat[b]{\includegraphics[width=3.75in]{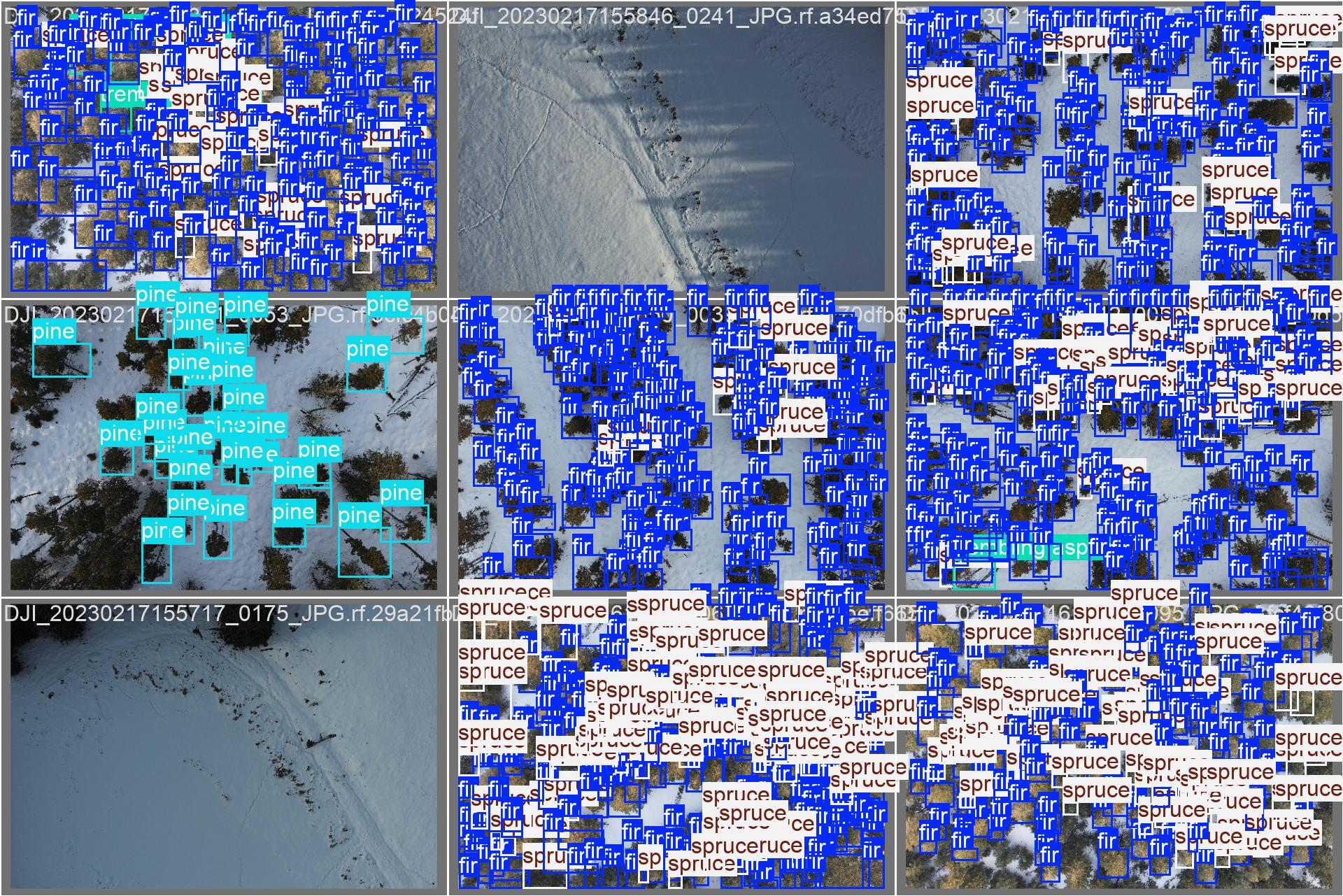}}
\caption{The classification result of YOLOv11 from our DW}
\label{train_val}
\end{figure*}

\begin{figure*}[!t]
\centering
\includegraphics[width=7in]{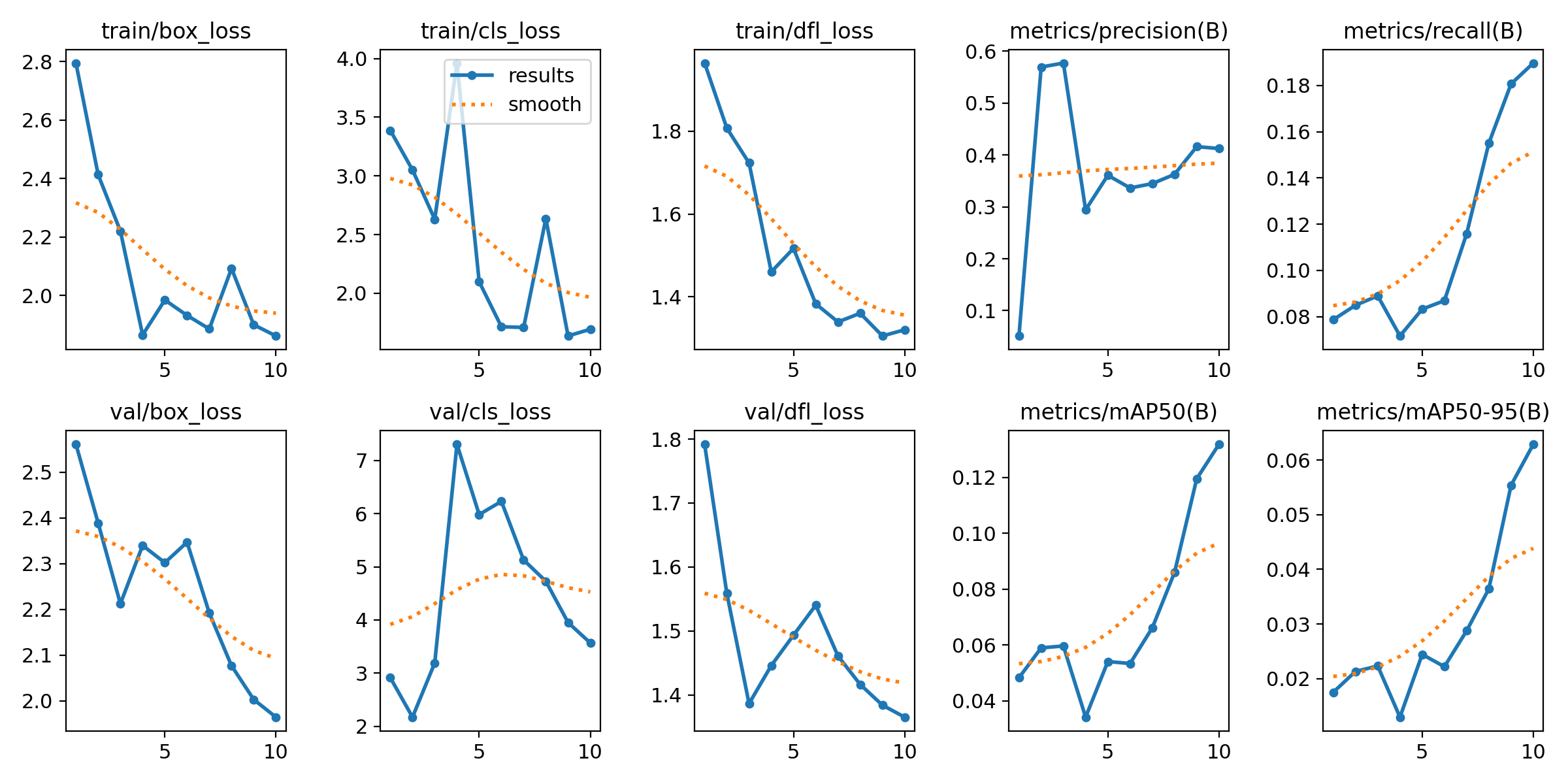}
\caption{The classification precision and recall from YOLOv11}
\label{results}
\end{figure*}

\section{Utilizing Digital Research Infrastructure for Scalability}
It evident by the estimations that as our data will grow, we will require more computing resources. The Digital Research Infrastructure (DRI) provided by the Digital Research Alliance of Canada (DRAC) allows for large scaled data processing, storage and analysis for forest inventory management and image processing. For research data storage and management, DRI provides access to various resources, including scalable and secure cloud based storage and Research Data Management (RDM) Assistant. Researchers can these to store and retrieve large datasets seamlessly while also complying with data security protocols and best practices.\\
The DRI makes it easier for multiple institutions to collaborate by providing shared access to computational resources and data repositories. Additionally, they provide access to open source software through their High Performance Computing (HPC) clusters, which are essential for creating an easy to access environment for performing machine learning and data analysis operations using our data. These HPC clusters also include access to Graphical Processing Units (GPUs) for accelerated performance of our workflows.\\
For this year's Resource Allocation Competition we have applied for cloud resources including 60TB storage, which is to be divided among test and development versions of the DW. Also, we have requested HPC resources including 460 cpu core years, with 4 processor cores and 8GB memory per core and 26 GPU years of the H100 3g40GB GPU model. These resource requirement were calculated based on our estimations for data scaling.\\
The support provided by DRAC is instrumental for equitable access to computing resources for research and promoting interdisciplinary research among institutions.

\section{Future Works}
Work to continue this research involves expanding the data schema to include dimension tables for video, text, and remote sensors. These additional dimensions will enhance the DW's ability to facilitate the integration and analyses of multimodal data, such as video feeds for real-time detection, textual data for contextual information, and sensor readings for environmental metrics. This integration will provide a more comprehensive and robust ITS classification and open opportunities for further research in the effective combination of these diverse data sources.. 

\section{CONCLUSION}
The integration of YOLO into a DW designed to identify and classify ITS offers significant benefits for urban and forestry planning and management. The design ensures optimized performance and storage efficiency by applying a star schema with three dimension tables (Date, Species, and Image) and a fact table for Tree Metrics. The YOLO framework streamlines data ingestion by extracting and classifying relevant tree data, reducing storage overhead while providing actionable insights. 

This DW architecture is well suited for diverse use cases, from monitoring urban tree health to managing forest ecosystems. Its scalability enables the seamless addition of future dimension tables for video, text, and remote sensor data without disrupting existing operations. As a result, the system can evolve alongside emerging technologies and growing data sources, making it a robust and resilient solution for environmental data management.  

The processes of analysis, forest inventories, and stand growth projection modeling generate extensive data, which is distributed among government and industry stakeholders to aid in informed decision-making. This emphasizes the necessity of a centralized information system.

\section*{Acknowledgement}
We extend our thanks and acknowledgements to the Department of Computer Science (CS) of Okanagan College (OC) and the University of Fraser Valley (UFV). Along with the OC and UFV Grant in Aid Committee \cite{acm_grants_guidelines}, for the funding and financial support related to the applied student research projects. 

Additionally, we would like to express our gratitude to both the OC and UFV research teams for their project info. Their information proved to be extremely useful with this phase of the research project.

Furthermore, we would like to thank Lucid Software Inc. \cite{lucidchart} for providing their platform, which facilitated our collaboration on diagrams and visualizations for this research.

In addition we also extend our appreciation to the Digital Resource Alliance of Canada, for providing resources related to this project.

\balance

\bibliographystyle{IEEEtran}

\bibliography{Kristina.bib}

\begin{thebibliography}{10}
\providecommand{\url}[1]{#1}
\csname url@samestyle\endcsname
\providecommand{\newblock}{\relax}
\providecommand{\bibinfo}[2]{#2}
\providecommand{\BIBentrySTDinterwordspacing}{\spaceskip=0pt\relax}
\providecommand{\BIBentryALTinterwordstretchfactor}{4}
\providecommand{\BIBentryALTinterwordspacing}{\spaceskip=\fontdimen2\font plus
\BIBentryALTinterwordstretchfactor\fontdimen3\font minus \fontdimen4\font\relax}
\providecommand{\BIBforeignlanguage}[2]{{%
\expandafter\ifx\csname l@#1\endcsname\relax
\typeout{** WARNING: IEEEtran.bst: No hyphenation pattern has been}%
\typeout{** loaded for the language `#1'. Using the pattern for}%
\typeout{** the default language instead.}%
\else
\language=\csname l@#1\endcsname
\fi
#2}}
\providecommand{\BIBdecl}{\relax}
\BIBdecl

\bibitem{wang2024individual}
\BIBentryALTinterwordspacing
L.~Wang, D.~Lu, L.~Xu, D.~T. Robinson, W.~Tan, Q.~Xie, H.~Guan, M.~A. Chapman, and J.~Li, ``Individual tree species classification using low-density airborne multispectral lidar data via attribute-aware cross-branch transformer,'' \emph{Remote Sensing of Environment}, vol. 315, p. 114456, 2024. [Online]. Available: \url{https://www.sciencedirect.com/science/article/pii/S0034425724004826}
\BIBentrySTDinterwordspacing

\bibitem{chen2021new}
\BIBentryALTinterwordspacing
C.~Chen, L.~Jing, H.~Li, and Y.~Tang, ``A new individual tree species classification method based on the resu-net model,'' \emph{Forests}, vol.~12, no.~9, p. 1202, 2021. [Online]. Available: \url{https://www.mdpi.com/1999-4907/12/9/1202}
\BIBentrySTDinterwordspacing

\bibitem{qin2022individual}
H.~Qin, W.~Zhou, Y.~Yao, and W.~Wang, ``Individual tree segmentation and tree species classification in subtropical broadleaf forests using uav-based lidar, hyperspectral, and ultrahigh-resolution rgb data,'' \emph{Remote Sensing of Environment}, vol. 280, p. 113143, 2022.

\bibitem{white2016remote}
J.~C. White, N.~C. Coops, M.~A. Wulder, M.~Vastaranta, T.~Hilker, and P.~Tompalski, ``Remote sensing technologies for enhancing forest inventories: A review,'' \emph{Canadian Journal of Remote Sensing}, vol.~42, no.~5, pp. 619--641, 2016.

\bibitem{lausch2018monitoring}
A.~Lausch and P.~J. Leitão, \emph{Monitoring Vegetation Diversity and Health through Spectral Traits and Trait Variations Based on Hyperspectral Remote Sensing}.\hskip 1em plus 0.5em minus 0.4em\relax CRC Press, 2018.

\bibitem{kimball2013data}
R.~Kimball and M.~Ross, \emph{The Data Warehouse Toolkit: The Definitive Guide to Dimensional Modeling}.\hskip 1em plus 0.5em minus 0.4em\relax John Wiley \& Sons, 2013.

\bibitem{farhadi2018yolov3}
A.~Farhadi and J.~Redmon, ``Yolov3: An incremental improvement,'' in \emph{Computer vision and pattern recognition}, vol. 1804.\hskip 1em plus 0.5em minus 0.4em\relax Springer Berlin/Heidelberg, Germany, 2018, pp. 1--6.

\bibitem{redmon2017yolo9000}
J.~Redmon and A.~Farhadi, ``Yolo9000: better, faster, stronger,'' in \emph{Proceedings of the IEEE conference on computer vision and pattern recognition}, 2017, pp. 7263--7271.

\bibitem{yolo11_ultralytics}
\BIBentryALTinterwordspacing
G.~Jocher and J.~Qiu, ``Ultralytics yolo11,'' 2024. [Online]. Available: \url{https://github.com/ultralytics/ultralytics}
\BIBentrySTDinterwordspacing

\bibitem{khan2023data}
Z.~Khan, ``Data mining and data warehouse: An in-depth review,'' \emph{ResearchGate}, 12 2023.

\bibitem{parihar2024algorithmic}
A.~Parihar, D.~Sareen, D.~Huitema, K.~Cormier, Y.~Khmelevsky, G.~Hains, and A.~Wong, ``Algorithmic trading machine learning modelling and forecasting subsystems integration and data transformation process automation for a data warehouse,'' \emph{Authorea Preprints}, 2024.

\bibitem{ghadimi2025databay}
\BIBentryALTinterwordspacing
M.~Ghadimi, N.~Baghayi, and A.~Shateri, ``Databay: A unified platform for automating data warehouse management, real-time data processing, and ensuring data quality and monitoring,'' \emph{TechRxiv}, 01 2025. [Online]. Available: \url{https://www.techrxiv.org/doi/full/10.36227/techrxiv.173834980.04708005/v1}
\BIBentrySTDinterwordspacing

\bibitem{bc_forest_inventory}
\BIBentryALTinterwordspacing
{Government of British Columbia}, ``Forest inventory,'' \textit{n.d.}, accessed: 2025-01-29. [Online]. Available: \url{https://www2.gov.bc.ca/gov/content/industry/forestry/managing-our-forest-resources/forest-inventory}
\BIBentrySTDinterwordspacing

\bibitem{roboflow_uav_tree}
\BIBentryALTinterwordspacing
{Roboflow}, ``Uav tree identification dataset,'' \textit{n.d.}, accessed: 2025-01-29. [Online]. Available: \url{https://universe.roboflow.com/arura-uav/uav-tree-identification-new/browse?queryText=&pageSize=50&startingIndex=0&browseQuery=true}
\BIBentrySTDinterwordspacing

\bibitem{nrcan_forest_carbon}
\BIBentryALTinterwordspacing
{Natural Resources Canada}, ``Forest carbon,'' \textit{n.d.}, accessed: 2025-01-29. [Online]. Available: \url{https://natural-resources.canada.ca/climate-change/climate-change-impacts-forests/forest-carbon/13085}
\BIBentrySTDinterwordspacing

\bibitem{acm_grants_guidelines}
\BIBentryALTinterwordspacing
{ACM Okanagan}, ``Grants-in-aid fund guidelines,'' \textit{n.d.}, accessed: 2025-01-29. [Online]. Available: \url{https://acm-assets.okanagan.bc.ca/digitalassetredirect/asset/did/5008/}
\BIBentrySTDinterwordspacing

\bibitem{lucidchart}
\BIBentryALTinterwordspacing
{Lucidchart}, ``Lucidchart - online diagramming and visual solution software,'' \textit{n.d.}, accessed: 2025-01-29. [Online]. Available: \url{https://www.lucidchart.com/}
\BIBentrySTDinterwordspacing

\end{thebibliography}

\end{document}